\documentstyle[aps,epsf]{revtex}
\def\tr{\,{\rm tr}\,}
\def\eqnn#1{(\ref{#1})}

\def\cmp{{ \sl Comm. Math. Phys. }}

\def\npb{{ \sl Nucl. Phys. }}
\def\prc{{ \sl Phys. Rep. }}
\def\prd{{ \sl Phys. Rev. }}
\def\prl{{ \sl Phys. Rev. Lett. }}
\def\plb{{ \sl Phys. Lett. }}

\def\dslash{\hbox{$\partial$\kern-0.5em\raise0.3ex\hbox{/}}}
\def\pslash{\hbox{{\it p}\kern-0.48em\raise-0.3ex\hbox{/}}}
\def\half{{1\over2}}
\def\tmu{\tilde \mu}
\def\gfive{\gamma_5}
\def\figno#1{{\it fig. \ref{fig:#1}}}
\def\intx{\int_0^1\!\!\!dx\,}
\def\intpx{\int_0^1\!\!\!dx\,}
\def\inty{\int_0^1\!\!dy\,}
\def\intk{\int\!\!{d^2k\over(2\pi)^2}\,}
\def\sqp{\sqrt{1-4M^2/p^2}}
\def\sqpp{\sqrt{4M^2/p^2-1}}
\def\wf#1#2{\tilde\varphi_{#1#2}}
\def\wfp#1#2{\tilde\psi_{#1#2}}
\def\vp#1{\varphi^{(#1)}}
\def\vpp#1{\varphi^{\prime(#1)}}
\def\hvp{\hat\varphi}
\def\gam#1{b_{#1}}
\def\coup{{a^2 N\over 4\pi}}
\def\coupp{{a_5^2 N\over 4\pi}}
\def\pint{P\!\!\int}
\def\bb#1{B(#1,#1)}
\draft
\begin{document}
\title{
  Physics of the gauged four fermi model in 
  $(1+1)$ dimensions 
}
\author{Kenichiro Aoki\footnote{E--mail: {\tt ken@phys-h.keio.ac.jp}}
and Kenji Ito\footnote{E--mail: {\tt kito@th.phys.titech.ac.jp}}
  }
\address{$^a$Hiyoshi Dept. of Physics, Keio University, {\it
    4--1--1} Hiyoshi, Kouhoku--ku, Yokohama 223--8521, Japan\\
  $^b$Dept. of Physics, Tokyo Institute of Technology,
  2--12--1 Oh-okayama, Meguro-ku, Tokyo {\it 152--0033}, Japan}
\date{\today }
\maketitle

\begin{abstract}
  We analyze a two dimensional model of gauged fermions with
  quartic couplings in the large--$N$ limit.  This combines the
  {}'t~ Hooft model and the Gross--Neveu model where the
  coupling constants of both theories are arbitrary.  Analytic
  equations describing the meson states of the theory are
  derived and are solved systematically using various methods.
  The physics of the model is investigated.
\end{abstract}
\section{Introduction}
\label{sec:intro}
Solvable models have greatly contributed to our understanding of
the dynamics of quantum field theories.  Two such
$(1+1)$--dimensional models solvable in the large $N$ limit
which are ``classics'' in this regard, are the {}'t~Hooft
model\cite{thooft}, gauge theory with fermionic matter, and the
Gross--Neveu model\cite{gn}, a model with a four fermi coupling.
(For reviews on the subject, see, for instance, \cite{revs}.)
Both these models are ``solvable'' from first principles yet
they are {\it not} integrable models in the usual sense, except
for the case of massless Gross--Neveu model \cite{integrable}.
These types of models are hard to come by and we believe that
they contribute to the understanding of more realistic theories.
Indeed, the {}'t~Hooft model and the Gross--Neveu model have a
wide area of applicability, as evidenced by the contribution of
these models in many areas of physics, including but not
restricted to particle and nuclear physics, even recently.  In
this paper, we solve a model in the large--$N$ limit wherein the
four fermi couplings and the gauge coupling coexist with
arbitrary strengths, thereby enlarging this class of models in
an essential way.  We extend and generalize the work of Burkardt
\cite{burkardt}, who derived the meson bound state equation in
the gauged four fermi model and analyzed the equation from a
somewhat different approach from ours.

The contents of this paper are as follows: First, in section 2,
we solve the Gross--Neveu model using the Bethe--Salpeter
equation, which, to our knowledge, has not been presented
previously.  In this section, we fix the notation and summarize
the physics of the Gross--Neveu model, which will be useful
later on.  In section 3, we analyze the theory of gauged
fermions with  four fermi couplings in the large--$N$ limit.
The model combines and generalizes the model of {}'t~Hooft and
the model of Gross and Neveu.  The model is more general than
the Gross--Neveu model even when the gauge coupling is zero.
Next we derive analytic equations for the meson states of the
theory.  Renormalization prescription is given and equations
involving only finite physical parameters are derived.  In
section 4, methods are presented in detail for solving the meson
state equations systematically.  Using these methods, we derive
various results on the physical properties of the model which
are analyzed in section 5.  We end with a summary and a more
general discussion regarding the model in Section 6.
\section{The massive Gross--Neveu model}
\label{sec:gn-model}
In this section, we analyze the Gross--Neveu model and derive
the Bethe--Salpeter equation for the fermion--antifermion
(meson) channel in a spirit similar to that of {}'t ~Hooft's
analysis of two dimensional QCD \cite{thooft}.  This method is
distinct from the methods applied to the Gross--Neveu model
previously\cite{gn,revs,itakura}\ and will provide useful
parallels for some aspects of the model to be discussed below.
We compare the results obtained from the Bethe--Salpeter
equations to those obtained from the usual approach and also
summarize the physics of the model pertinent for the sequel.

The Gross--Neveu model we analyze in this section has the
following Lagrangian;
\begin{equation}
  \label{gn-lag}
  {\cal L}= \overline\psi^j\left(i\dslash-m\right)\psi_j
  + {a^2\over2} \left[\left(\overline\psi^j\psi_j\right)^2
      -\left(\overline\psi^j\gamma_5\psi_j\right)^2\right]
    \qquad j=1,2\ldots,N
\end{equation}
This is equivalent to the following Lagrangian written using the 
auxiliary real scalar fields $\sigma,\chi$,
\begin{equation}
  \label{gn-lag-aux}
  {\cal L}= \overline\psi^j i\dslash\psi_j
  -\half \left(\sigma^2+\chi^2\right) + 
  a \overline\psi^j\left(\sigma+i\chi\gfive\right)\psi_j
  -{m\over a}\sigma
    \qquad j=1,2\ldots,N
\end{equation}

\subsection{The equation for the ``meson'' bound states}
\label{sec:gn-bs}
We will use light cone coordinates
$v^\pm=v_\mp\equiv1/\sqrt2\left(v^0\pm v^1\right)$ below.  In
light cone coordinates, the gamma matrices become simple in the
chiral basis
\begin{equation}
  \label{gammas}
  \gamma^+=\pmatrix{0 & 0 \cr \sqrt2 & 0 \cr}\qquad
  \gamma^-=\pmatrix{0 & \sqrt2 \cr 0 & 0 \cr}\qquad
  \gamma_5=\pmatrix{1 & 0 \cr 0 & -1 \cr}
\end{equation}
$1/N$ expansion is performed by expanding in powers of $1/N$
while treating $a^2N$ as a quantity of order one.  We use the
metric $(+,-)$.

{}From the interactions in the Lagrangian \eqnn{gn-lag}\ we
obtain a self consistent equation for the propagator in the
large $N$ limit.  As is clear from the interactions, these
corrections do {\it not} have any momentum dependence.
Therefore, it can be effectively summarized in a mass parameter,
which we shall call $M$.  The fermions in this model are
physical particles and the parameter $M$ is the physical mass of
these fermions, which will be determined self consistently.

\begin{figure}[htp]
  \begin{center}
    \leavevmode
    \epsfxsize=\hsize\epsfbox{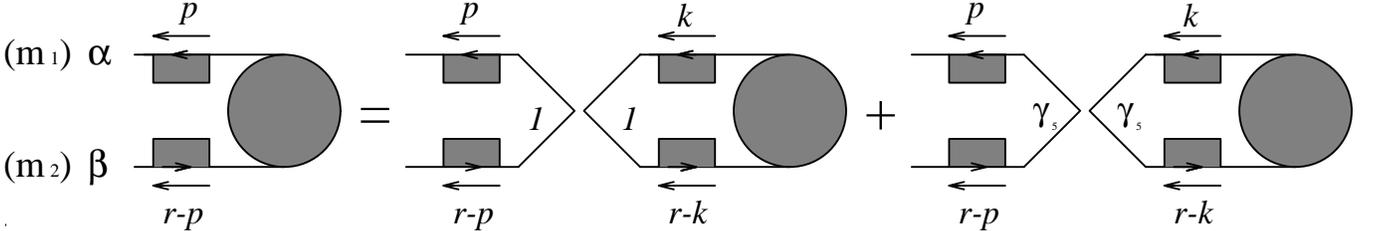}
    \caption{Bound state equation for the fermion--antifermion
      system in the Gross--Neveu model} 
    \label{fig:gn-bs}
  \end{center}
\end{figure}

Using the full propagator, we may straightforwardly obtain the
Bethe--Salpeter equation for what we shall call the ``meson
wavefunction'', $\wfp\alpha\beta$, from the contributions
graphically represented in \figno{gn-bs}\ as:
\begin{equation}
  \label{gn-bs}
  \wfp\alpha\beta(p,r) =  2 a_b^2N \left[
  S_{\alpha1}(p)S_{1\beta}(p-r)\intk\wfp22(k,r) 
  + S_{\alpha2}(p)S_{2\beta}(p-r)\intk\wfp11(k,r)\right]
\end{equation}
$S(p)$ is the full fermion propagator in this model, which is
none other than the tree--level propagator with the mass $m$
replaced by $M$.  $a_b$ denotes the {\it bare} four fermi
coupling.  Following 't Hooft, we integrate over one component
of the momentum $p_+$ and define $\wf\alpha\beta\equiv\int
dp_+\wfp\alpha\beta$.  Then, after some computation, the
equations for the meson wavefunction $\wfp\alpha\beta$ reduce to
the following equations for the components:
\begin{eqnarray}
  \label{gn-bs-components}
  {2\pi\over a_b^2N}\left[\tmu^2-{1\over x(1-x)}\right]\wf11(x)
  &=& 
  -{1\over2(1-x)}\left[\tmu^2-{1\over x}+{1\over 1-x}\right]
  \inty\wf11(y)   
  \nonumber\\ &&\qquad
  +{1\over x(1-x)}\inty\wf22(y) 
  \nonumber\\
  {2\pi\over a_b^2N}\left[\tmu^2-{1\over x(1-x)}\right]\wf22(x)
  &=& 
  -{1\over2x}\left[\tmu^2+{1\over x}-{1\over 1-x}\right]
  \inty\wf22(y)   
  \nonumber\\ &&\qquad
  +{1\over x(1-x)}\inty\wf11(y)
  \nonumber\\
  {2\pi\over a_b^2N}{M\over\sqrt2 r_-}
  \left[\tmu^2-{1\over x(1-x)}\right]\wf12(x)
  &=& 
  {1\over1-x}\inty\wf11(y)-{1\over x}\inty\wf22(y)
  \nonumber\\
  {2\pi\over a_b^2N}{\sqrt2 r_-\over M}
  \left[\tmu^2-{1\over x(1-x)}\right]\wf21(x)
  &=& 
  -{1\over2x(1-x)}\left[\tmu^2-{1\over x}+{1\over 1-x}\right]
  \inty\wf11(y) 
  \nonumber\\
  &&\qquad+{1\over2x(1-x)}\left[\tmu^2+{1\over x}-{1\over 1-x}\right]
  \inty\wf22(y) 
\end{eqnarray}
Here, we defined the momentum fraction of the incoming
antifermion $x\equiv p_-/r_-$ and the mass squared of the bound
state in units of the fermion mass squared as
$\tmu^2=2r_+r_-/M^2$.  Without any loss of generality, we may
define $\int dx\wf11=1,\ \int dx\wf22=C$, where $C$ is to be
determined later.  Then all the meson wavefunctions
$\wf\alpha\beta$ may be solved algebraically using the equations
\eqnn{gn-bs-components} as follows.  The consistency with the
normalization of $\wf11$ requires that
\begin{equation}
  \label{gn-raw}
  {4\pi\over a_b^2 N} = \int _0^1\!dx\,
  {{\tmu^2\over2}-2C-2+{1\over2x(1-x)}\over 1-\tmu^2 x(1-x)} 
\end{equation}
The compatibility of this with the normalization condition for
$\wf22$ requires that $C=\mp1$.  These two cases correspond to
the meson states $\chi$ and $\sigma$ respectively.  We obtain
the equations determining the masses of $\chi$ and $\sigma$ as
\begin{equation}
  \label{gn-sigma-pi}
  \chi:  \qquad {4\pi\over a^2N} =
  \intx {\tmu^2_\chi\over 1-\tmu_\chi^2 x(1-x)}
  ,\qquad
  \sigma:  \qquad {4\pi\over a^2N} =
  \intx {\tmu^2_\sigma-4\over 1-\tmu_\sigma^2 x(1-x)}
\end{equation}
Here, we renormalized the coupling constant as 
\begin{equation}
  \label{gn-renormalization}
  {4\pi\over a^2N}= {4\pi\over a_b^2N} -   \half\intpx
  {1\over x(1-x)}  
\end{equation}
The integral needs to be regularized at the endpoints $0,1$
which is not explicitly expressed here.  The same
renormalization was employed in the light front Hamiltonian
approach in \cite{burkardt}.  This regularization is effectively
an ultraviolet cutoff, which will become clear below.  The meson
wavefunctions for $\chi,\sigma$ can also be obtained
algebraically as
\begin{equation}
  \label{gn-wf}
  \chi:\qquad \wf12^\chi(x)= {\rm const.}\times{1\over
    1-\tmu_\chi^2  x(1-x)}
  ,\qquad
  \sigma: \qquad \wf12^\sigma(x)= {\rm const.}\times{1-2x\over 
    1-\tmu_\sigma^2  x(1-x)}
\end{equation}
The component $\wf12$ is shown here since it corresponds to the
relevant component of the meson wavefunction in the {}'t~Hooft
model \cite{thooft}\ and will also be the essential component in
our analysis of the gauged four fermi model.  The wavefunction
for $\chi$ is consistent with the results obtained using light
front Hamiltonian methods \cite{itakura,burkardt}.  Other
components of the wavefunction can also be computed
algebraically using \eqnn{gn-bs-components}.
\subsection{The analysis of the Gross--Neveu model using scalars 
  and its relation to the Bethe--Salpeter equation}
\label{sec:gn-pot}
In this section, we clarify the relation between the results
obtained above using the Bethe--Salpeter equation and the
results obtained from using the auxiliary scalar fields
$\sigma,\chi$ as in the Lagrangian \eqnn{gn-lag-aux}. Since the
approach of using scalars is more standard and is explained
elsewhere, we refer the derivation to the literature
\cite{gn,revs}.

The full propagators for the $\sigma,\chi$ fields are \cite{gn}
\begin{eqnarray}
  \label{gn-propagators}
  \sigma:\qquad D^{-1}_\sigma(p^2) &=& 
  1+{a_b^2 N\over2\pi}  \left[\ln {M^2\over\Lambda^2} +
    B(p^2,M^2)\right] \nonumber\\
  \chi:\qquad D^{-1}_\chi(p^2) &=& 
  1+{a_b^2 N\over2\pi}  \left[\ln {M^2\over\Lambda^2} +
    {1\over1-4M^2/p^2}B(p^2,M^2)\right]
\end{eqnarray}
where the $a_b$ is the bare coupling and $\Lambda$ is the
ultraviolet momentum scale cutoff.  (There is a mild abuse of
notation here; this bare coupling is in principle not the same
as the one in the previous section since the regularization
methods are different.) The function $B(p^2,M^2)$ is defined as
\begin{equation}
  \label{b-def}
  B(p^2,M^2)\equiv\sqp\ln{\sqp+1\over\sqp-1}
  = 2\sqpp\cot^{-1}\sqpp
\end{equation}
The renormalized coupling $a$ defined in
\eqnn{gn-renormalization}\ is related to the bare coupling $a_b$
here as
\begin{equation}
  \label{gn-renorm}
 {2\pi\over a^2N}= {2\pi\over a_b^2N}+ \ln {M^2\over\Lambda^2} 
\end{equation}
We find that the equations for the poles in the propagators for
$\sigma,\chi$ in \eqnn{gn-propagators}\ indeed agree with the
equations \eqnn{gn-sigma-pi} in this renormalization scheme.
The propagators have cuts for $p^2>4M^2$ corresponding to the
production of physical fermion--antifermion pair of mass $M$
each.

The effective potential for the scalars may be obtained by
computing the contributions from the fermion loops in the
large--$N$ limit as 
\begin{equation}
  \label{gn-pot}
  V(\sigma,\chi) = 
    \half\left(\sigma^2+\chi^2\right) - {m\over a_b}\sigma 
    + {a_b^2N\over4\pi}\left(\sigma^2+\chi^2\right)
    \left(\ln {a_b^2\left(\sigma^2+\chi^2\right)\over \Lambda^2} 
      -1 \right)
\end{equation}
By minimizing the potential, we obtain a vacuum expectation
value $\langle\sigma\rangle$ for $\sigma$.  The physical mass is
related to the vacuum expectation value simply as
$M=a_b\langle\sigma\rangle$, without loss of generality.  We
obtain a relation between the bare and the renormalized
parameters
\begin{equation}
  \label{mm-relation}
  {M\over a^2N}={m\over a_b^2N}
\end{equation}
Since the mass is dynamically generated in the Gross--Neveu
model even when $m=0$, the chiral limit corresponds to
$a^2N\rightarrow\infty$. 
\subsection{Physics of the Gross--Neveu model}
\label{sec:gn-physics}
\begin{figure}[htb]
  \begin{center}
    \leavevmode
    \epsfxsize=12cm\epsfbox{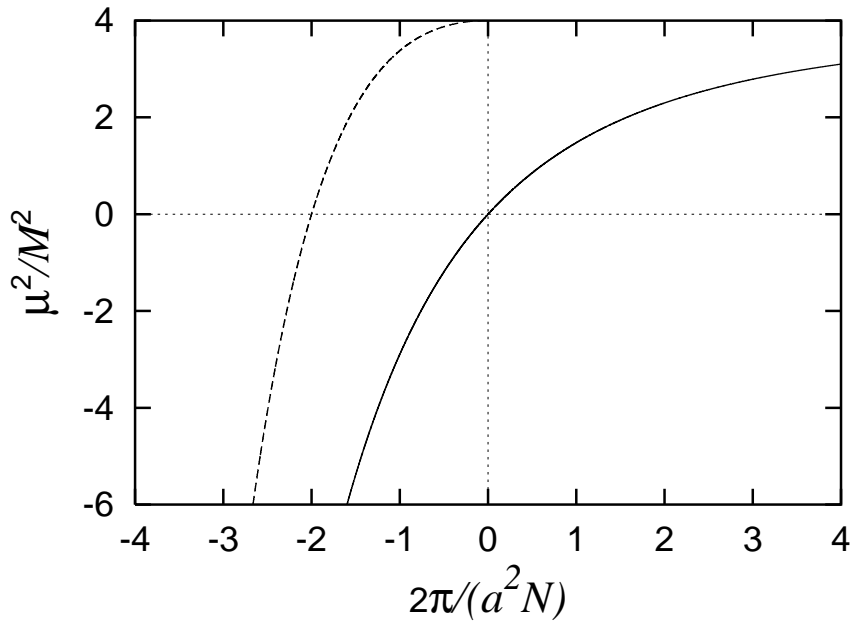}
    \vskip0.5cm
    \caption{Mass squared of $\chi$ (solid) and $\sigma$
    (dashed) in units of $M^2$ versus the inverse     coupling
    $2\pi/(a^2N)$.}
    \label{fig:gn-spectrum}
  \end{center}
\end{figure}

\begin{figure}[htb]
  \begin{center}
    \leavevmode
    \epsfxsize=\hsize\epsfbox{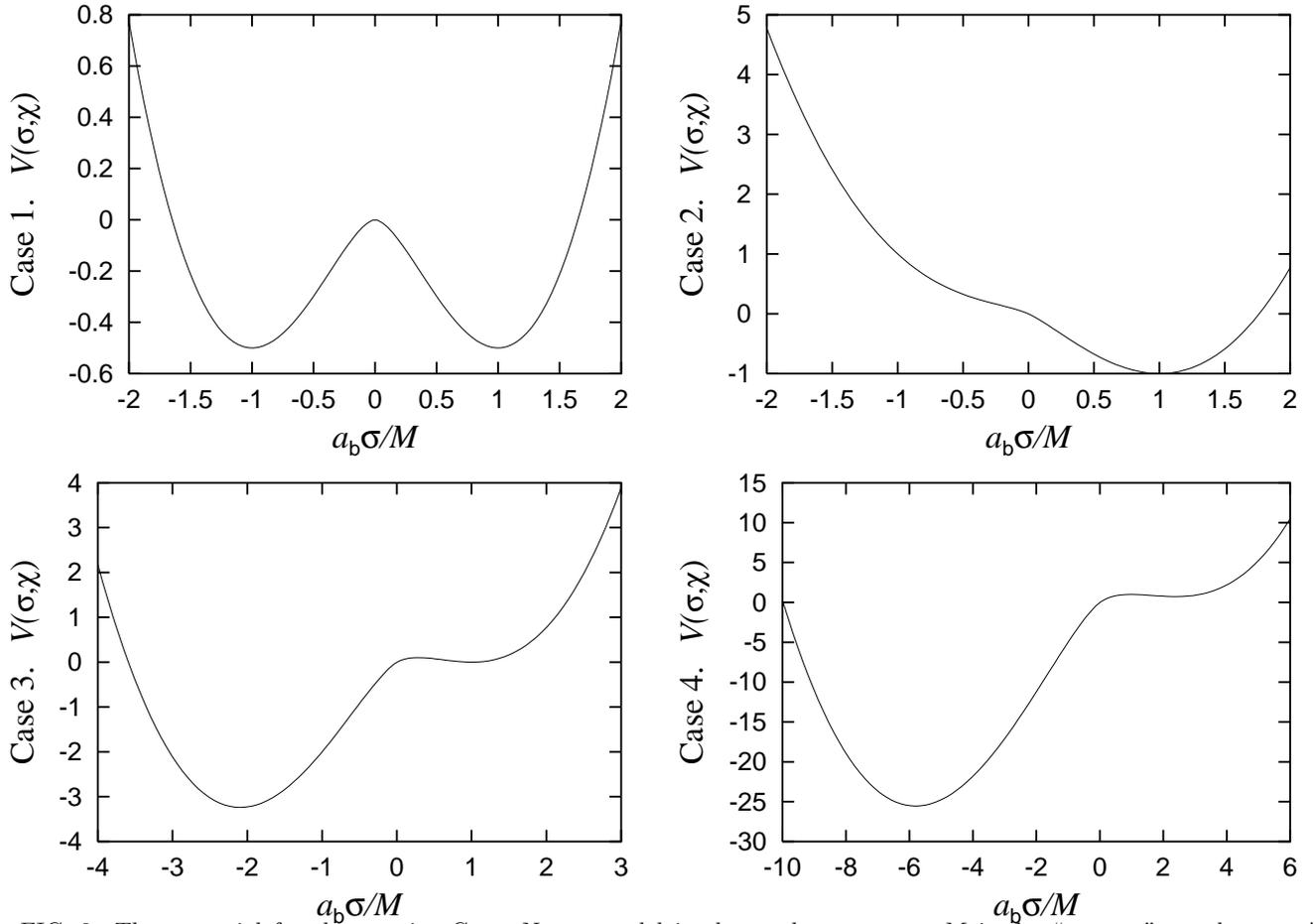}
    \caption{The potential for the massive Gross--Neveu model in 
      the $\sigma$--plane. $a_b\sigma=M$ is the ``vacuum''
      we choose. An example is given for the each of the four
      cases 1---4 explained in the text.  The values chosen are
      $(2\pi)/a^2N=0,1,-1$ and $-3$, respectively.
      Vertical scale is in units of $NM^2/2\pi$.} 
    \label{fig:gn-pot}
  \end{center}
\end{figure}

Here, we briefly summarize the physics of the Gross--Neveu
model, in particular, emphasizing the salient features and its
underlying physics which will be useful later on.  The
Lagrangian\ \eqnn{gn-lag}\ has two parameters, $m$ and $a$.  Due
to dimensional transmutation, the model is determined
essentially by only one parameter.  We can solve the equations
\eqnn{gn-sigma-pi} or the pole equations of the propagators
\eqnn{gn-propagators} to obtain the masses of the scalars
$\sigma,\chi$.  We plot the spectrum of $\sigma,\chi$ against
the inverse coupling in \figno{gn-spectrum}.
We understand the various regions in the coupling constant
as follows:
\begin{enumerate}
\item $1/a^2N=0$: The chiral point.  Here, the masses for
  $\sigma$ and $\chi$ are respectively $2M$ and zero.  The
  wavefunction for $\chi$, $\wf12^\chi(x)$, is a constant in
  this limit, similarly to the {}'t~Hooft model.  $\chi$ is the
  ``Nambu--Goldstone'' boson of the theory.  Strictly speaking,
  Nambu--Goldstone boson does not exist in $(1+1)$ dimensions
  \cite{ngb}, yet it is well known that many physical aspects of
  the higher dimensional theories are also seen in $(1+1)$
  dimensional theories, especially in the large $N$ limit.  A
  similar massless boson bound state exists in the {}'t~Hooft
  model in the chiral limit.

  The status of the $\sigma$ particle is interesting; while the
  $\sigma$ particle is usually said to exist, its wavefunction
  $\wf12^\sigma(x)$ approaches const.$/(1-2x)$ and is singular
  in the limit $1/a^2N\rightarrow0$.  Physical decay into a
  fermion and an antifermion becomes kinematically allowed in
  this limit $\mu^2\rightarrow 4M^2$ and the singular behavior
  is due to this.  Similar behavior is also seen for $\chi$ in
  the limit $\mu^2\rightarrow 4M^2$.  The wavefunction is well
  behaved when $a^2N<0$, yet in this region, the vacuum is not
  physically stable, as explained below.
\item $a^2N>0$: This region is physically consistent.  The mass
  of $\chi$ is between zero and $2M$.  The meson wavefunction
  has a singular limit
  $\wf12^\chi\rightarrow$const.$\times(1-2x)^{-2}$ as
  $1/a^2N\rightarrow \infty$.  The pole in the $\sigma$
  propagator \eqnn{gn-sigma-pi}\ that exists for $a^2N\leq0$
  ceases to exist in this regime and there is no bound state
  corresponding to $\sigma$.  Poles corresponding to resonance
  states also do not exist, unlike the Gross--Neveu model in
  $(2+1)$--dimensions\cite{revs}.
\item 
  $a^2N\leq-\pi$: 
  While $\sigma$ scalar has a  finite mass, $\chi$ is tachyonic.
  From the potential, we may understand this as follows;  we are 
  at an unstable vacuum where the potential is locally a minimum 
  in the $\sigma$ direction yet {\it maximal} in the $\chi$
  direction.   Choosing the physically sensible vacuum reduces
  this case to the previous physically consistent case. 
\item $-\pi<a^2N<0$: We again have chosen an unstable
  vacuum. this vacuum is unstable in both $\sigma$ and $\chi$
  directions so that both $\sigma,\chi$ are tachyonic.  Had the
  correct vacuum been chosen, the theory reduces to the $a^2N>0$
  case above.
\end{enumerate}
We plot the potential for these various cases in \figno{gn-pot}\
along the plane $\chi=0$.

\section{The gauged four fermi model}
\label{sec:gnh}
In this section, we analyze the gauged four fermi model.  We
obtain the Bethe--Salpeter equation for the fermion--antifermion
bound states --- which we shall call the ``mesons'' for obvious
reasons --- and perform the renormalization of the model.  We
further systematically solve the Bethe--Salpeter equation to
obtain the masses and the wavefunctions of the meson states. 

The Lagrangian of the gauged four fermi model we work with is
\begin{equation}
  \label{tgn-lag}
  {\cal L} =-\frac{1}{2}\tr\left(F_{{\mu}{\nu}}F^{{\mu}{\nu}} \right)
  +\sum_f{\overline{\psi}}_f(i
  {\hbox{{\it D}\kern-0.52em\raise0.3ex\hbox{/}}}-m_f){\psi}_{f}
  +\frac{a^2}{2}\sum_{(f,f')}({\overline{\psi_{f'}}}{\psi_f})
  ({\overline{\psi_f}}{\psi_{f'}})
  -\frac{a_5^2}{2}\sum_{(f,f')}({\overline{\psi_{f'}}}{\gamma}_5{\psi_f})
  ({\overline{\psi_{f}}}{\gamma}_5{\psi_{f'}})
  \qquad
\end{equation}
The covariant derivative is defined as
$D_\mu\equiv\partial_\mu-igA_\mu$, where $g$ is the gauge
coupling constant.  The color indices have been suppressed.
Index $f$ denotes the ``flavor'' index and is included here
since we shall consider bound states involving fermions of
different masses.  The motivations for such a generalization is
obvious when we want to apply the model to more realistic
situations. The Lagrangian generically respects the global
flavor symmetry $\left[U(1)_V\right]^N$, which is enlarged to
$SU(N)_V\times U(1)_V$ when all the fermion masses $m_f$ are the
same.  This symmetry is further enlarged to the chiral flavor
symmetry $SU(N)_L\times U(1)_L\times SU(N)_R\times U(1)_R$ when
all $m_f=0$.  The Gross--Neveu model we dealt with in the
previous section corresponds to the case when there is no gauge
coupling, only a single flavor type and $a^2=a_5^2$.
\subsection{The equations describing the mesons}
Below, we will derive Bethe--Salpeter equations for the fermion
bound states by using methods similar to those of {}'t~Hooft
\cite{thooft}. The situation is intrinsically more complicated
than that of {}'t~Hooft since the four fermi interactions mix
the fermions of different chirality and the Bethe--Salpeter
equation can no longer be straightforwardly reduced to a one
dimensional equation.  We fix the gauge to the light-cone gauge
$A_-=A^+=0$.  Light-cone gauge has the advantage that there are
no gluon self-interactions in $(1+1)$ dimensions.  We take the
large $N$ limit by letting $N$ go to infinity keeping
$g^2N,a^2N, a_5^2N$ fixed.

\begin{figure}[htp]
  \begin{center}
    \leavevmode
    \epsfxsize=140mm\epsfbox{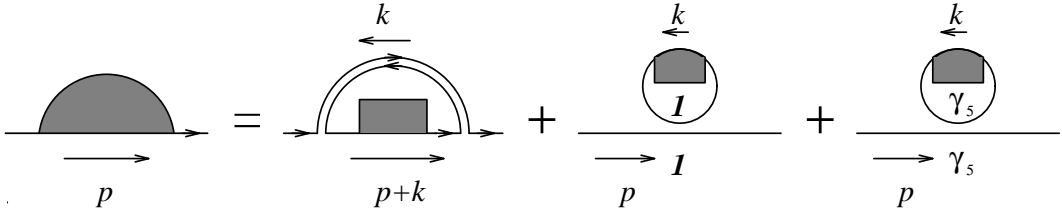}
    \caption{The self consistent equation for the propagator in
      the gauged four fermi model} 
    \label{fig:tgn-sd}
  \end{center}
\end{figure}

First, we obtain the full propagator self consistently from the
Schwinger--Dyson equation.  Diagrammatically, the equation may
be expressed as in \figno{tgn-sd} in the large $N$ limit.
Solving the equation we obtain the full propagator $S(p;M_f)$ as
\begin{equation}
  \label{tgn-prop}
  S^{-1}(p;M_f) =
  -i\left[\pslash-M_f+i\epsilon+
    \frac{g^2N}{2\pi}\left(\frac{sgn(p_-)}{{\lambda}_-}
      -\frac{1}{p_-}\right)\gamma^+\right]
\end{equation}
where $M_f$ is the mass parameter containing the quantum
corrections, as in the previous section.  We introduced a cutoff
${\lambda}_-$ for the infrared divergence in $p_-$ integral.

\begin{figure}[htp]
  \begin{center}
    \leavevmode
    \epsfxsize=\hsize\epsfbox{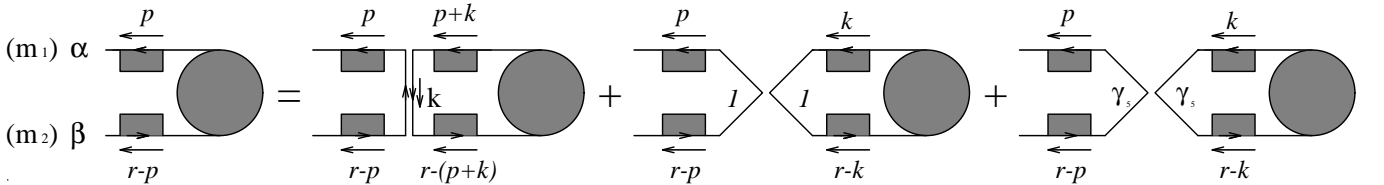}
    \caption{The bound state equation for the fermion--antifermion
      system in the gauged four fermi model.}
    \label{fig:tgn-bs}
  \end{center}
\end{figure}

The bound state equation for fermion--antifermion bound state
may be derived in the large $N$ limit, which is
diagrammatically depicted in \figno{tgn-bs}.
Denoting the wavefunction of the bound state as
${\psi}_{{\alpha}{\beta}}(p,r)$, the Bethe--Salpeter equation
can be derived as the following matrix equation in the large $N$
limit
\begin{eqnarray}
  {\psi}(p,r)&=&
  -i2g^2N\, S(p;M_1)\gamma^+  \intk
  {\psi}(p+k,r)\gamma^+ S(p-r;M_2)
  {1\over k_-^2}  
  \nonumber \\   &-&
  i{a_b^2N}\,  S(p;M_1)S(p-r;M_2)\intk\tr{\psi}(k,r)
  \nonumber \\   &+&
  ia_{5b}^2N
  S(p;M_1)\gamma_5S(p-r;M_2)
  \intk\tr\left(\gamma_5{\psi}(k,r)\right)
\label{tgn-bs} 
\end{eqnarray}
The suffix $b$ on the couplings indicates that these couplings
are bare parameters.  We define, as in the previous section,
$
{\varphi}_{{\alpha}{\beta}}(p_-,r){\equiv}\int_{-\infty}^{\infty}dp_+
{\psi}_{{\alpha}{\beta}}(p,r)$.
Then the bound state equations may be obtained for the components
as
\begin{eqnarray}
  \label{tgn11}
  &&\left({\mu}^2-\frac{2r_-}{{\lambda_-}}-\frac{{\beta}_1-1}{x} -
    \frac{{\beta}_2-1}{1-x}\right){\varphi}_{11}(x)
  \nonumber\\ 
  &&\quad=
  \frac{M_2}{\sqrt2r_-(1-x)}
  \left(\frac{2r_-}{ {\lambda_-}}{\varphi}_{12}(x)+
    \pint_0^1dy\frac{{\varphi}_{12}(y)}{(x-y)^2}\right)
  \nonumber \\   &&\qquad
  +\frac{a_b^2N}{2{\pi}}
  \frac{1}{4x(1-x)}\left[2\sqrt{{\beta}_1{\beta}_2}-x
    \left(\mu^2-\frac{2r_-}{\lambda_-}-\frac{\beta_1-1}{x} -
      \frac{-{\beta}_2-1}{1-x}\right)\right]
  \int_0^1dy [{\varphi}_{11}(y)+{\varphi}_{22}(y)] 
  \nonumber \\ 
  &&\qquad-
  \frac{a_{5b}^2N}{2{\pi}}
  \frac{1}{4x(1-x)}\left[2\sqrt{{\beta}_1{\beta}_2}+x
    \left({\mu}^2-\frac{2r_-}{{\lambda_-}}-\frac{{\beta}_1-1}{x} -
      \frac{-{\beta}_2-1}{1-x}\right)\right]
  \int_0^1dy  [{\varphi}_{11}(y)-{\varphi}_{22}(y)]
\end{eqnarray}
\begin{eqnarray}
  \label{tgn22}
  &&
  \left({\mu}^2-\frac{2r_-}{\lambda_-}-\frac{\beta_1-1}{x} -
      \frac{{\beta}_2-1}{1-x}\right){\varphi}_{22}(x)
  \nonumber  \\   &&\quad
  =-\frac{M_1}{\sqrt2r_-x}\left(\frac{2r_-}{\lambda_-}
    {\varphi}_{12}(x)+P\inty\frac{{\varphi}_{12}(y)}{(x-y)^2}\right)
  \nonumber \\   &&\qquad
  +\frac{a_b^2N}{2{\pi}}
  \frac{1}{4x(1-x)}\left[2\sqrt{{\beta}_1{\beta}_2}-(1-x)
    \left({\mu}^2-\frac{2r_-}{\lambda_-}-\frac{-{\beta}_1-1}{x} -
      \frac{{\beta}_2-1}{1-x}\right)\right] 
    \inty [{\varphi}_{11}(y)+{\varphi}_{22}(y)] 
    \nonumber \\   &&\qquad
    +\frac{a_{5b}^2N}{2{\pi}}
    \frac{1}{4x(1-x)}\left[2\sqrt{{\beta}_1{\beta}_2}+(1-x)
    \left({\mu}^2-\frac{2r_-}{\lambda_-}-\frac{-{\beta}_1-1}{x} -
      \frac{{\beta}_2-1}{1-x}\right)\right] 
  \inty[{\varphi}_{11}(y)-{\varphi}_{22}(y)]
\end{eqnarray}
\begin{eqnarray}
  \label{tgn12}
  &&
  \left({\mu}^2-\frac{{\beta}_1-1}{x} -
      \frac{{\beta}_2-1}{1-x}\right){\varphi}_{12}(x)\nonumber\\
  &&\qquad =- \pint_0^1dy\frac{{\varphi}_{12}(y)}{(x-y)^2}
  -\frac{M_1(1-x)-M_2x}{2\sqrt2x(1-x)}\frac{a_b^2r_-}{g^2}\int_0^1dy
  \left[{\varphi}_{11}(y)+{\varphi}_{22}(y)\right]
  \nonumber \\   &&\qquad
  +\frac{M_1(1-x)+M_2x}{2\sqrt2x(1-x)}
  \frac{a_{5b}^2r_-}{g^2}\int_0^1dy
  \left[{\varphi}_{11}(y)-{\varphi}_{22}(y)\right]
\end{eqnarray}
\begin{eqnarray}
  \label{tgn21}
  &&\left({\mu}^2-\frac{2r_-}{\lambda_-}-\frac{{\beta}_1-1}{x} 
    -\frac{{\beta}_2-1}{1-x}\right){\varphi}_{21}(x)
  \nonumber\\ &&\qquad
  =     \frac{M_1M_2}{2r_-^2 x(1-x)}\left(\frac{2r_-}{\lambda_-}
    {\varphi}_{12}(x)+\pint_0^1dy\frac{{\varphi}_{12}(y)}{(x-y)^2}\right)
  \nonumber \\ &&\qquad
  +  \frac{a_b^2N}{2{\pi}}\frac{1}{4\sqrt2x(1-x)}
  \biggl[
    - \left({\mu}^2-\frac{2r_-}{{\lambda_-}}-\frac{{\beta}_1-1}{x} -
      \frac{-{\beta}_2-1}{1-x}\right){M_1\over r_-}
    \nonumber \\   &&\qquad\quad +
     \left({\mu}^2-\frac{2r_-}{\lambda_-}-\frac{-\beta_1-1}{x} 
      -      \frac{{\beta}_2-1}{1-x}\right){M_2\over r_-}
  \biggr]
  \inty \left[{\varphi}_{11}(y)+{\varphi}_{22}(y)\right] 
  \nonumber \\   &&\qquad 
  -
  \frac{a_{5b}^2N}{2{\pi}}\frac{1}{4\sqrt2x(1-x)}
  \biggl[
    \left({\mu}^2-\frac{2r_-}{{\lambda_-}}-\frac{{\beta}_1-1}{x} -
      \frac{-{\beta}_2-1}{1-x}\right){M_1\over r_-}
  \nonumber \\   &&\qquad\quad +
     \left({\mu}^2- \frac{2r_-}{{\lambda_-}}-\frac{-{\beta}_1-1}{x} -
      \frac{{\beta}_2-1}{1-x}\right){M_2\over r_-}
  \biggr]
  \inty\left[{\varphi}_{11}(y)-{\varphi}_{22}(y)\right] 
\end{eqnarray}
Here we defined 
\begin{equation}
  \label{xy-def}
  \beta_i\equiv \pi {M^2_i\over g^2N},\quad
   \mu^2\equiv 
   {\pi 2 r_+ r_- \over g^2N},\quad
  x\equiv\frac{p_-}{r_-},\quad
  y\equiv\frac{k_-}{r_-}
\end{equation}
and $\pint$ denotes the principal part integral.  When the four
fermi couplings $a^2,a_5^2$ are absent, the equation
\eqnn{tgn12}\ is the {}'t~Hooft equation, which is a closed
equation by itself.  Here, all the equations are coupled and
they need to be disentangled in a more sophisticated manner.

Superficially, we have as many equation as the unknowns --- the
meson wavefunctions, $\varphi_{\alpha\beta}$'s.  However, we
expect the infrared cutoff $\lambda_-$ to decouple from these
physical equations, so that these equations are possibly
over--constrained.  It may be shown that all these equations
consistently reduce to the following equations
\eqnn{tgn11f}---\eqnn{tgn21f}\ and \eqnn{tgn12}.  These
equations do {\it not} involve the infrared cutoff but are yet
to be renormalized:
\begin{eqnarray}
  \label{tgn11f}
  {\varphi}_{11}(x)=\frac{-M_2}{\sqrt2r_-(1-x)}{\varphi}_{12}(x)
  &-&\frac{1}{4(1-x)}\frac{a_b^2N}{2\pi}\int_0^1dy
  [{\varphi}_{11}(y)+{\varphi}_{22}(y)] \nonumber \\ 
  &-&\frac{1}{4(1-x)}\frac{a_{5b}^2N}{2\pi}\int_0^1dy
  [{\varphi}_{11}(y)-{\varphi}_{22}(y)] 
\end{eqnarray}
\begin{equation}
  \label{tgn22f}
  {\varphi}_{22}(x)=\frac{M_1}{\sqrt2r_-x}{\varphi}_{12}(x)
  -\frac{a_b^2N}{2\pi}\frac{1}{4x}\inty
  [{\varphi}_{11}(y)+{\varphi}_{22}(y)] 
  +\frac{a_{5b}^2N}{2\pi}\frac{1}{4x}\inty
  [{\varphi}_{11}(y)-{\varphi}_{22}(y)]
\end{equation}
\begin{eqnarray}
  \label{tgn21f}
  {\varphi}_{21}(x)=-\frac{M_1M_2}{2r_-^2x(1-x)}{\varphi}_{12}(x)
  &-&\frac{M_1-M_2}{4\sqrt2r_-x(1-x)}\frac{a_b^2N}{2\pi}\int_0^1dy
  [{\varphi}_{11}(y)+{\varphi}_{22}(y)] \nonumber \\ 
  &-&\frac{M_1+M_2}{4\sqrt2r_-x(1-x)}\frac{a_{5b}^2N}{2\pi}\int_0^1dy
  [{\varphi}_{11}(y)-{\varphi}_{22}(y)] 
\end{eqnarray}
{}From these equations we may derive the following closed
equation for $\varphi_{12}(\equiv\varphi)$, whose suffix we
shall omit for brevity from now on.
\begin{eqnarray}
  \label{tgnf}
  {\mu}^2{\varphi}(x) &\equiv& 
  H\varphi(x)
  \nonumber \\
  &=&   \left(\frac{{\beta}_1-1}{x}
    +\frac{{\beta}_2-1}{1-x}\right){\varphi}(x) -
  P\int_0^1dy\frac{{\varphi}(y)}{(x-y)^2} 
  \nonumber \\   &&\quad
  -\frac{\sqrt{{\beta}_1}(1-x)-{\sqrt
      {{\beta}_2}}x}{x(1-x)}\frac{\int_0^1dy
    \frac{{\sqrt{{\beta}_1}}(1-y)-{\sqrt{{\beta}_2}}y}
    {y(1-y)}{\varphi}(y)}{\frac{4\pi}{a_b^2N}
    +\frac{1}{2}\int_0^1\frac{dx}{x(1-x)}}
  \nonumber \\ &&\quad
  -\frac{\sqrt{{\beta}_1}(1-x)+{\sqrt
      {{\beta}_2}}x}{x(1-x)}\frac{\int_0^1dy
    \frac{{\sqrt{{\beta}_1}}(1-y)
      +{\sqrt{{\beta}_2}}y}{y(1-y)}{\varphi}(y)}
  {\frac{4\pi}{a_{5b}^2N}+\frac{1}{2}\int_0^1\frac{dx}{x(1-x)}}
\end{eqnarray}
We shall often refer to $H$ as the Hamiltonian.  It is clear
that this equation reduces to the {}'t~ Hooft equation when
$a^2=a_5^2=0$ and that it reduces to the Gross--Neveu model case
obtained in the previous section when $g=0$, $\beta_1=\beta_2$
and $a^2=a_5^2$.  Even when the gauge coupling is zero, this
model is more general than the massive Gross--Neveu model in
that it incorporates different four fermi couplings $a,a_5$ and
fermions of different masses.  This equation describes the
properties of the mesons in the gauged four fermi model.  When
the fermion masses are equal,
${\beta}_1={\beta}_2({\equiv}{\beta})$, the equation takes a
substantially simpler form;
\begin{eqnarray}
  \label{tgnf-eqmass}
  {\mu}^2{\varphi}(x) &=&\frac{{\beta}-1}{x(1-x)}{\varphi}(x) -
  P\int_0^1dy\frac{{\varphi}(y)}{(x-y)^2} 
  \nonumber \\ &&\quad
  -\frac{{\beta}}{x(1-x)}\frac{\int_0^1dy
    \frac{{\varphi}(y)}{y(1-y)}}{\frac{4\pi}{a_b^2N}
    +\frac{1}{2}\int_0^1\frac{dx}{x(1-x)}}
  -\frac{{\beta}(1-2x)}{x(1-x)}\frac{\int_0^1dy
    \frac{1-2y}{y(1-y)}{\varphi}(y)}
  {\frac{4\pi}{a_{5b}^2N}+\frac{1}{2}\int_0^1\frac{dx}{x(1-x)}}
\end{eqnarray}

When the fermion masses are equal, $\beta_1=\beta_2$, {\it and }
when the couplings are equal, $a^2=a_5^2$, the bound state
equation for the mesons in the gauged four fermi model
\eqnn{tgnf}\ reduces to the equation given by Burkardt
\cite{burkardt}.  Burkardt obtained a renormalized form of the
equation when the meson wavefunction is an even function of the
momentum fraction by using an operator identity involving the
divergence of the axial current.  In contrast, below, we will
renormalize the more general bound state equation \eqnn{tgnf}\ 
and reduce the equations to its renormalized form without using
any further identities.
\subsection{Renormalization of the gauged four fermi model}
\label{sec:tgn-renormalization}
The equations derived in the previous section
\eqnn{tgn11f}--\eqnn{tgnf} are expressed in terms of bare
quantities.  The equation for the meson wavefunction should be
expressible in terms of renormalized quantities and the
Hamiltonian matrix elements between physical states should be
finite.  From these conditions, we may derive the
renormalization for the couplings and the boundary conditions
for the meson wavefunction.  The coupling constants are
renormalized as follows
\begin{equation}
  \label{tgn-renormalization}
  {4\pi\over a^2N}= {4\pi\over a_b^2N} 
  - \half\intx {1\over x(1-x)},\qquad
  {4\pi\over a_5^2N}= {4\pi\over a_{5b}^2N} - 
  \half\intx {1\over x(1-x)}
\end{equation}
As in the Gross--Neveu model, these integrals have been
regularized which is not explicitly denoted here.  It should be
noted here that this generalizes the renormalization of the
coupling constant in the Gross--Neveu model
\eqnn{gn-renormalization}. The mass parameter $M$ needs no
renormalization.  This is again consistent with the
renormalization in both the Gross--Neveu model and the
{}'t~Hooft model.  We expand the meson wavefunction as 
\begin{equation}
  \label{meson-expand}
  \varphi(x)= \vp0 + \vp1 (1-2x) + \hvp(x)
\end{equation}
where $\vp0,\vp1$ are constants and
$\hvp(x)/\left[x(1-x)\right]$ is integrable at $x=0,1$.  Then,
the boundary conditions for the meson wavefunction are
\begin{equation}
  \label{tgn-bc}
  \pmatrix{ \gam+ & (1+4\coup)\gam- \cr 
    \gam-  & (1+4\coupp)\gam+\cr}
  \pmatrix{\vp0\cr \vp1\cr} = \intx{\hvp(x)\over x(1-x)}
  \pmatrix{ \coup & 0 \cr  0  & \coupp\cr}
  \pmatrix{ \gam+ & \gam- \cr  \gam-  & \gam+\cr}
  \pmatrix{1 \cr 1-2x\cr}
\end{equation}
Here, we defined $ \gam\pm\equiv\left( \sqrt\beta_1\pm
  \sqrt\beta_2\right)/2$.  In particular, when the coupling
constants are equal, $a^2=a_5^2$, {\it or} when the masses are
equal, $\beta_1=\beta_2$, the boundary conditions simplify to
\begin{equation}
  \label{tgn-bc-simple}
  \pmatrix{\vp0\cr \vp1\cr} = \intx{\hvp(x)\over x(1-x)}
  \pmatrix{\coup \cr {\coupp(1-2x)\over 1+4\coupp}\cr}
\end{equation}
The meson wavefunction does {\it not} vanish at the boundaries.
This property is similar to that of the Gross--Neveu model but
{\it unlike} that of the {}'t~Hooft model.  When the
Gross--Neveu couplings are zero, the wavefunction {\it does}
vanish at the boundaries, thereby recovering the boundary
conditions of {}'t~Hooft. Also, it is instructive to check that
the equation for the meson wavefunction \eqnn{tgnf}\ and the
boundary conditions \eqnn{tgn-bc-simple}\ for $\beta_1=\beta_2$
reduce exactly to the equations \eqnn{gn-sigma-pi}\ in the
Gross--Neveu model for $\sigma$ and $\chi$ scalars when $\vp0=0$
and $\vp1=0$, respectively.

The equation for the meson states is now reduced to 
\begin{eqnarray}
  \label{finite-meson-eq}
  H\varphi(x) &=& \mu^2\varphi(x)\nonumber\\
  &=&
  \left({\beta_1-1\over x}+{\beta_2-1\over
      1-x}\right)  \hvp(x)
  -\pint_0^1\!\!dy\,{\hvp(y)\over (y-x)^2}
  +2\vp1\left( -\beta_1+\beta_2 + \ln{1-x\over x}\right)
\end{eqnarray}
subject to the boundary conditions \eqnn{tgn-bc}.  It is
straightforward to check that the Hamiltonian is Hermitean under
the given boundary condition.  The {\it explicit} dependence on
the coupling constants is contained in the non--trivial boundary
conditions.  The problem has been reduced to that of solving a
well defined integral equation.
Below, we restrict to the case of scalar and pseudo scalar
couplings being equal, $a=a_5$, for simplicity.  We will,
however, consider fermions of different masses in general.  The
more general case may be dealt with in a similar fashion,
involving somewhat more complicated formulas.  {}From here on,
we shall use the notation $G\equiv a^2N/(4\pi)$ to avoid
cluttering the formulas.  For the case $a=a_5$, the matrix
elements of the Hamiltonian may be simplified to the following
form which is appropriate for the application of variational
methods.
\begin{eqnarray}
  \label{matrix-elements-general}
  \left(\varphi',H\varphi\right)
  &=&
  {1\over 2G}\left(\beta_1+\beta_2\right)\overline{\vpp0}\vp0
  + \left[{1+4G\over 2G}
    \left(\beta_1+\beta_2\right)+2\right]\overline{\vpp1}\vp1
  \nonumber\\ &&\quad
  +2\intx\ln{1-x\over x}
  \left(\overline{\hvp'(x)}\vp1+\overline{\vpp1}\hvp(x)\right)
  \nonumber\\ &&\quad
  + \left({\beta_1+\beta_2\over2}-1\right)
  \intx{\overline{\hvp'(x)}\hvp(x)\over x(1-x)} 
  -\pint_0^1\int_0^1\!\!{dx\,dy\over(y-x)^2}\,
  \overline{\hvp'(x)}\hvp(y)
  \nonumber\\ &&\quad
  + \half\left(\beta_1-\beta_2\right)\biggl[
    {1\over G}
    \left(\overline{\vpp1}\vp0+\overline{\vpp0}\vp1\right)
    -4\intx\left(\overline{\hvp'(x)}\vp1
      +\overline{\vpp1}\hvp(x)\right)
    \nonumber\\ &&\qquad\quad
    +\intx{(1-2x)\overline{\hvp'(x)}\hvp(x)\over x(1-x)}   \biggr]
\end{eqnarray}
\section{Systematic methods for solving the meson bound state
  equation}
In this section, we show how the meson bound state equation
\eqnn{tgnf}\ may be solved systematically utilizing a finite
dimensional system of algebraic equations.  We will give
explicit formulas for two approaches, namely a variational
method using polynomials of the momentum fraction and a method
of working with the equation directly using a sinusoidal basis.
These methods will be used in the next section to investigate
some physical properties of the gauged four fermi model.

\subsection{Variational method}
\label{sec:variational}
Here, we shall use a variational method using polynomials of the
momentum fraction $x$ that satisfy the boundary condition
\eqnn{tgn-bc}.  In the {}'t~Hooft model, a similar method was
employed in \cite{power-ref}.  Without loss of generality, the
basis may be chosen to be
\begin{equation}
  \label{variational-basis}
  \varphi_{2k}(x)=G+{\left[x(1-x)\right]^k\over \bb k}
  ,\qquad
  \varphi_{2k+1}(x)=(1-2x)\left({G\over 1+4G}
  +{(2k+1)\left[x(1-x)\right]^k\over \bb k}\right)
  ,\qquad
  k=1,2,\ldots
\end{equation}
The normalization factor was chosen so as to make the matrix
elements be of order one.  The meson bound state equation
becomes
\begin{equation}
  \label{tgn-var-eigenproblem}
  \det({\mu}^2 N_{kl}-H_{kl})=0,\quad 
  H_{kl}\equiv\left(\varphi_k,H\varphi_l\right),\ 
  N_{kl}\equiv\left(\varphi_k,\varphi_l\right)  \qquad
  k,l=2,3,4{\ldots}
\end{equation}
In practice, we approximate the solution by using a finite
dimensional version of this equation.  The matrix elements can
be computed to be
\begin{eqnarray}
  \label{ham-elems}
  H_{2k,2l} &=& 
  \half\left(\beta_1+\beta_2-2\right) {\bb {k+l}\over \bb k \bb l}
  +{G\over2} \left(\beta_1+\beta_2\right)
  + {kl\over2(k+l)} \nonumber\\
  H_{2k+1,2l+1} &=& 
  \half\left(\beta_1+\beta_2-2\right) 
  {(2k+1)(2l+1)\bb {k+l}\over (2k+2l+1)\bb k \bb l}
  +{G\over2(1+4G)} \left(\beta_1+\beta_2\right)
  \nonumber\\  &&\quad
  +2\left({G\over1+4G}\right)^2
  +{G\over1+4G} \left({k\over k+1}+{l\over l+1}\right)
  +{kl(2k+1)(2l+1)\over  2(k+l)(k+l+1)}
  \nonumber\\
  H_{2k,2l+1} &=& H_{2l+1,2k} =
  \half\left(\beta_1-\beta_2\right)
  \left[{G\over 1+4G} {1\over 2k+1}
    + {(2l+1) \bb {k+l}\over (2k+2l+1) \bb k \bb l}
  \right]
\end{eqnarray}
\begin{eqnarray}
  \label{norm-elems}
  N_{2k,2l} &=& 
  {(k+l)\bb {k+l}\over 2(2k+2l+1)\bb k \bb l}
  + G^2 +G\left[{k\over 2(2k+1)}+{l\over 2(2l+1)}\right]
  \nonumber\\
  N_{2k+1,2l+1} &=& 
  {(2k+1)(2l+1)(k+l)\bb {k+l}\over 2(2k+2l+3)(2k+2l+1)\bb k \bb l}
  +{1\over3}\left({G\over1+4G}\right)^2 
  \nonumber\\&&\qquad
  + {G\over2(1+4G)}\left({k\over 2k+3}+{l\over 2l+3}\right)
  \nonumber\\
  N_{2k,2l+1}&=&N_{2k+1,2l} = 0
\end{eqnarray}
When the fermion masses are equal, $\beta_1=\beta_2$, the even and 
the odd sectors decouple.
\subsection{Multhopp's method}
\label{sec:multhopp}
Rather than using a variational method, we may choose to work
with the equation \eqnn{finite-meson-eq}\ directly.  By a clever
choice of basis, this singular integral equation may be brought
into an algebraic equation.  The method we use here generalizes
the methods used to numerically analyze the {}'t~Hooft equation
previously \cite{multhopp,ai}.    

We expand the meson wavefunction as 
\begin{equation}
  \label{sin-expand}
  \varphi(x) = \vp0-\vp1\cos\theta
  +\sum c_n\sin n\theta,\qquad 
  x\equiv \half(1+\cos\theta)
\end{equation}
$\vp{0,1}$ terms are absent in the {}'t~Hooft equation and can
be determined in our model from the boundary conditions
\eqnn{tgn-bc}. 
\begin{equation}
  \label{vp-sin}
  \vp0=2\pi G\sum_{n:\ \rm odd} c_n,\qquad
  \vp1 = -2\pi {G\over 1+4G} \sum_{n:\ \rm even}
  c_n,\qquad
\end{equation}
then the meson equation \eqnn{finite-meson-eq}\ becomes 
\begin{equation}
  \label{meson-eq-m-inf}
  \sum_n \left[\mu^2 \hat P_n(\theta) - \hat M_n(\theta)\right]
  c_n = 0
\end{equation}
where
\begin{eqnarray}
  \label{pmhat-def}
  \hat P_n(\theta) &\equiv&
  \sin n\theta+2\pi\cases{G& $n$: odd\cr 
    {G\over1+4G}\cos\theta& $n$: even\cr}\nonumber\\
  \hat M_n(\theta) &\equiv&
  2\left({\beta_1-1\over 1+\cos\theta}
    +{\beta_2-1\over 1-\cos\theta}\right)\sin n\theta
  +2\pi{n\sin n\theta\over\sin\theta}
    +\cases{0& $n$: odd\cr 
    4\pi{G\over1+4G}\left(\beta_1-\beta_2+
      \ln{1+\cos\theta\over1-\cos\theta}\right)
    & $n$: even\cr}
\end{eqnarray}

In practice, we truncate the basis to a finite dimensional one
and systematically analyze the convergence of the solution by
varying the dimension of this finite space, which we shall
denote by $K$.  Then the meson bound state equation may be reduced to
a generalized eigenvalue problem.
\begin{equation}
  \label{meson-eq-m}
  \sum_{n=1}^K \left[\mu^2 P_{mn}- M_{mn}\right] c_n = 0
\end{equation}
where
\begin{equation}
  \label{pm-def}
  P_{mn}\equiv\sum_{l=1}^K
  g_m(\theta_l)\hat P_n(\theta_l)
  ,\quad
  M_{mn}\equiv\sum_{l=1}^K
  g_m(\theta_l)\hat M_n(\theta_l)
  ,\qquad
  \theta_j\equiv\pi{j\over K+1}
\end{equation}
In this work, we choose $g_m(\theta)=2\sin m\theta/(K+1)$ as in
the {}'t~Hooft model. Other choices may be more convenient
depending on the parameters of the model.  Similarly to the
variational method given above, $M_{mn},P_{mn}\propto
\left(\beta_1-\beta_2\right)$ when $m-n\equiv1\ ({\rm mod}\ 2)$
and the even and odd sectors decouple completely when
$\beta_1=\beta_2$.  This property is preserved for any choice of
$g_m(\theta_l)$ as long as the property
$g_m(\theta_l)=(-1)^{m+1}g_m(\theta_{K+1-l})$ is
preserved. Unlike the variational method, however, the
approximate solution obtained by truncating to finite
dimensional solution space needs {\it not} be an upper bound on
the true solution and in general will not be.
\section{Physics of the gauged four fermi model}
\label{sec:physics}
First, we would like to determine the parameter region where the
behavior of the system is physically reasonable.  In particular,
the meson bound state should {\it not} be tachyonic.  Here, we
will perform the analysis for fermions of equal mass since we
expect tachyonic mesons in the flavor singlet channel {\it if}
tachyonic mesons exist at all.  Using the variational method for
the meson wavefunction in a manner similar to the previous
section, 
\begin{equation}
  \label{tgn-var}
  \varphi_\gamma(x) \equiv G+\bb \gamma^{-1}
  \left[x(1-x)\right]^\gamma\qquad\gamma>0
\end{equation}
we obtain an upper bound $\mu^2_\gamma$ for the meson mass
squared for each $\gamma>0$.  
\begin{equation}
  \label{tgn-upb}
  \mu^2_\gamma
  ={\left(\varphi_\gamma,H\varphi_\gamma\right)\over
    \left(\varphi_\gamma,\varphi_\gamma\right)}
  = {G\beta+{\gamma\over4}+(\beta-1)
    {\bb{2\gamma}\over \bb{\gamma}^2}    \over
    G^2+G{\gamma\over2\gamma+1}+{\gamma\over4\gamma+1}
    {\bb{2\gamma}\over \bb{\gamma}^2}}  \sim\cases{
    {\beta\over G} &$\gamma\sim0$\cr
    2\sqrt{2\pi\gamma}&$\gamma\sim\infty$\cr}
\end{equation}
This immediately establishes that when $\beta G <0$ tachyonic
mesons will appear.  While it is {\it logically} possible from
this analysis that the region with both $\beta<0$ and $G<0$ may
be physically consistent, it is unreasonable to expect so; in
practice, we find that tachyonic mesons appear for this case
also, when we have a large enough variational space.  Therefore,
we henceforth interest ourselves in the region $G>0$ and
$\beta>0$.

\begin{figure}[htbp]
  \begin{center}
    \leavevmode
    \epsfxsize=15cm\epsfbox{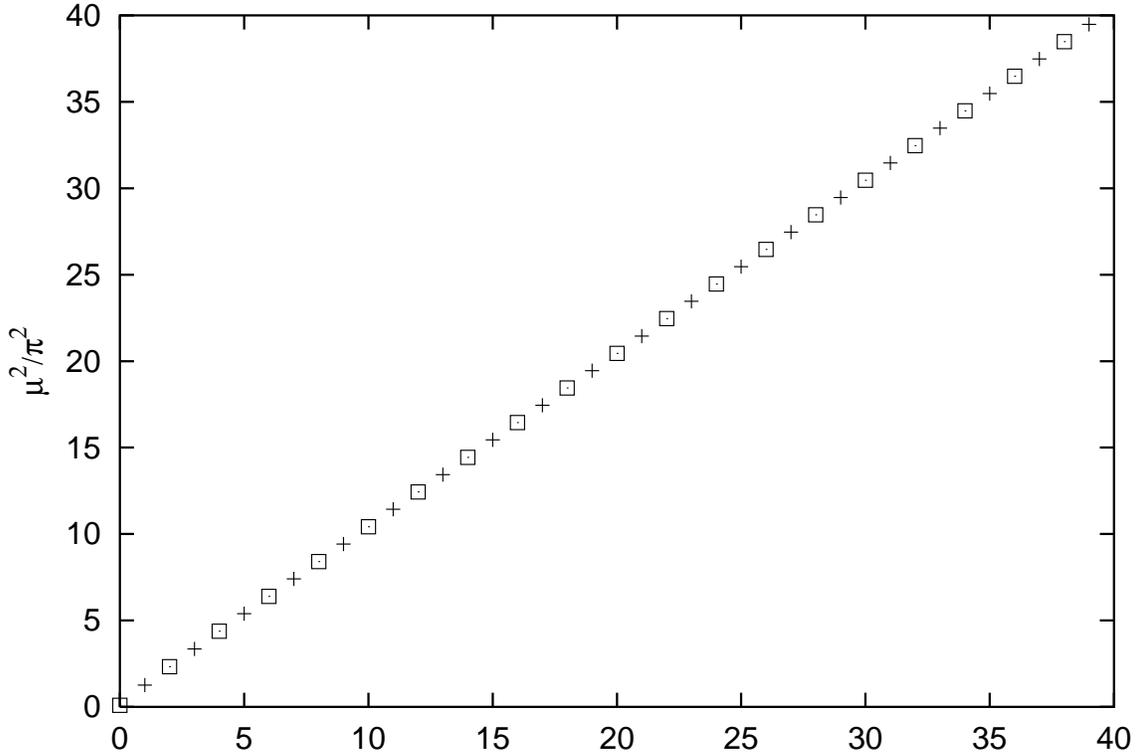}
    \caption{Mass squared normalized by
      $\pi^2$ of the meson states  versus the ``level number''
      for the case $\beta=1$, $G=1$. $\Box$ ($+$)'s denote states 
      whose wavefunctions are even (odd) under $x\leftrightarrow1-x$,}
    \label{fig:regge}
  \end{center}
\end{figure}

Using the methods explained in the previous section, we may
obtain the spectrum and the meson wavefunctions.  We plot the
spectrum and the wave functions for some typical cases below in
figures \figno{m-g}, \figno{m-b}\ and \figno{wf}.  We have added
a brief note on the convergence of the numerical data as an
appendix.
\begin{figure}[htbp]
  \begin{center}
    \leavevmode
    \epsfxsize=15cm\epsfbox{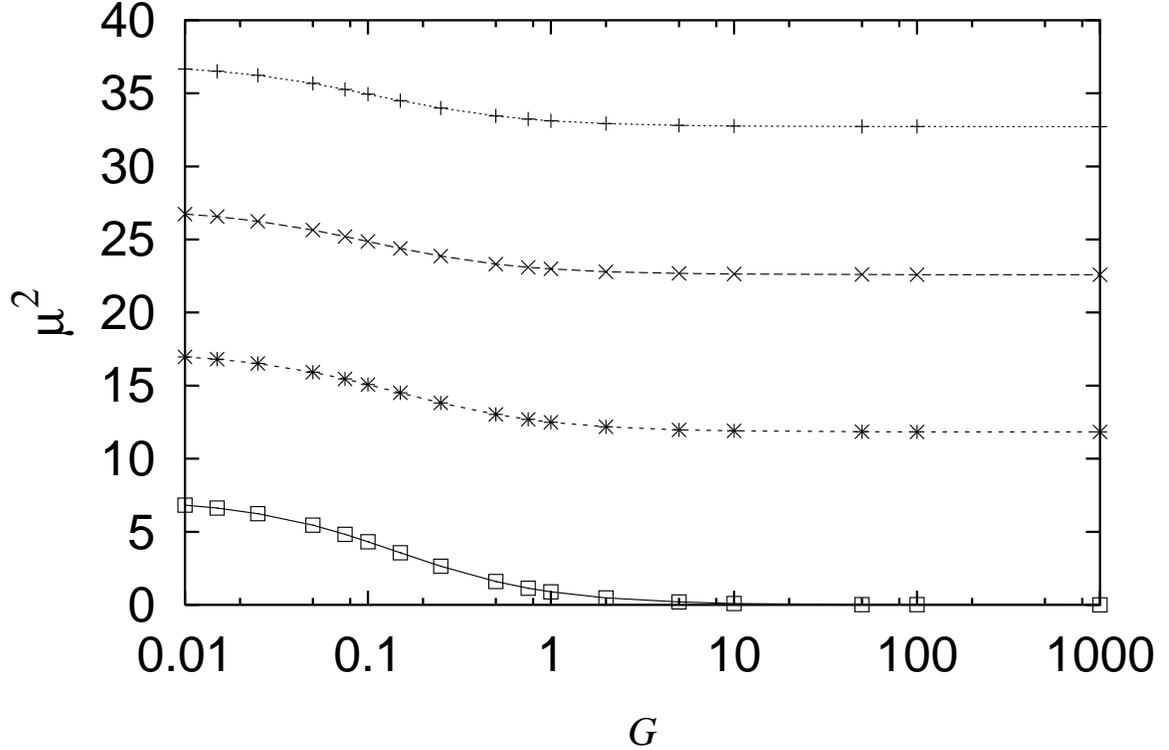}
    \caption{Mass squared of the lightest four meson states
      versus the coupling $G$ for $\beta=1$.} 
    \label{fig:m-g}
  \end{center}
\end{figure}
\begin{figure}[htbp]
  \begin{center}
    \leavevmode
    \epsfxsize=15cm\epsfbox{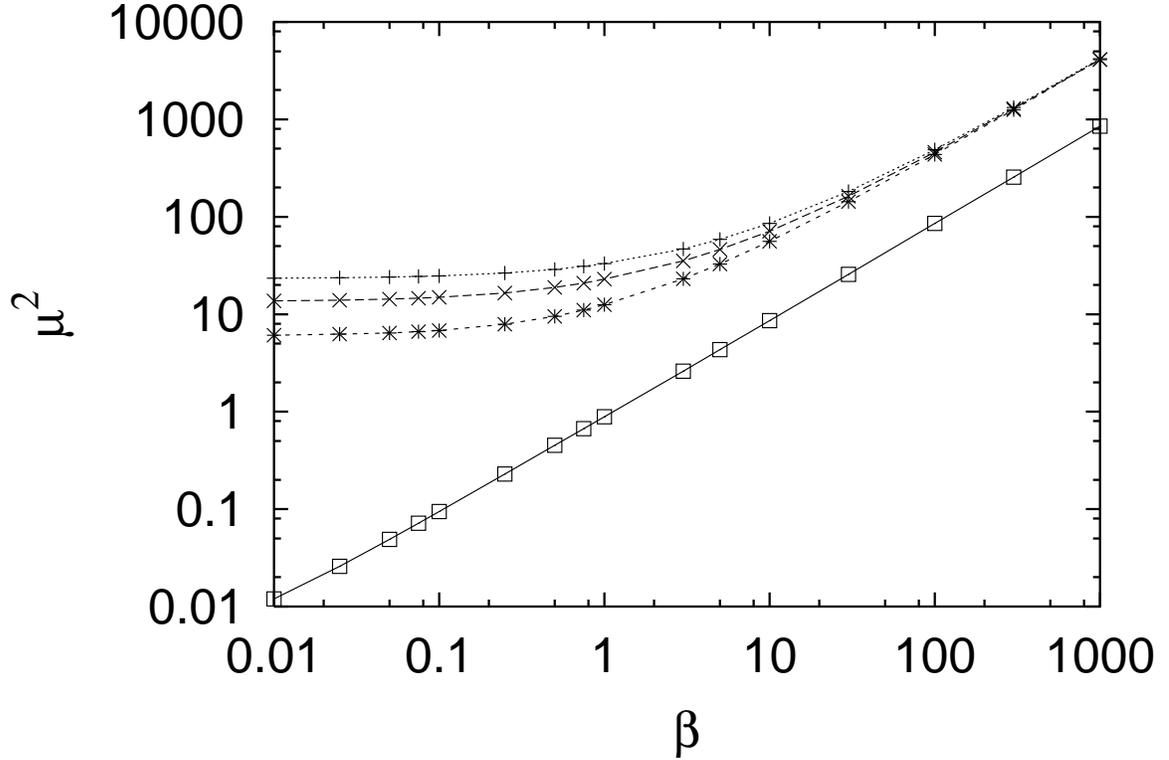}
    \caption{Mass squared of the lightest four meson states
      versus the fermion mass squared $\beta$ for the coupling
      $G=1$.} 
    \label{fig:m-b}
  \end{center}
\end{figure}
\begin{figure}[htb]
  \begin{center}
    \leavevmode
    \epsfxsize=\hsize\epsfbox{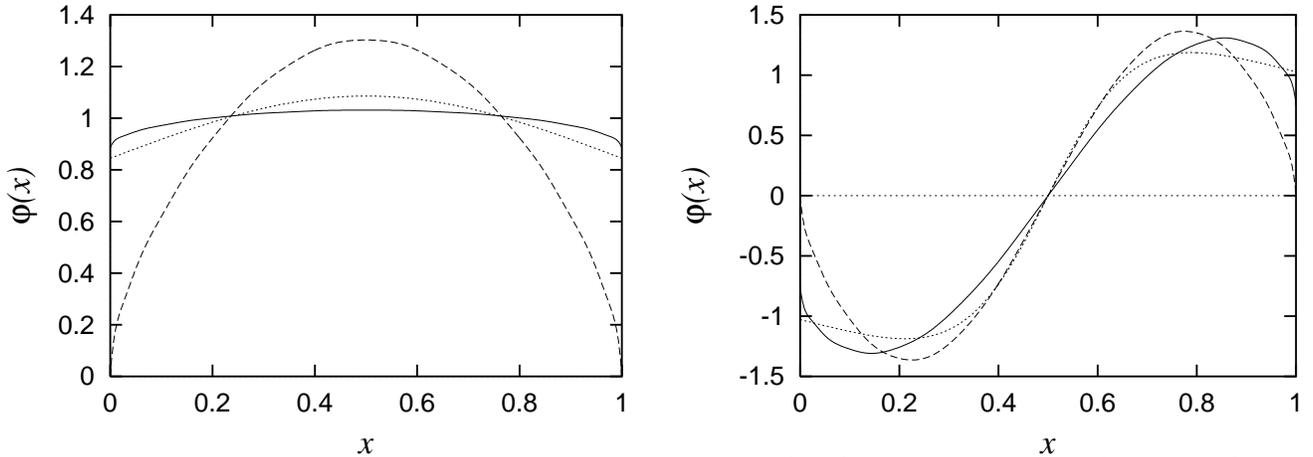}
    \caption{The meson wavefunctions for the gauged four fermi
    model (solid), the {}'t~Hooft model (long dashes) and the
    Gross--Neveu model (short dashes).  The functions
    are plotted for the lightest two meson states.  The
    parameters chosen for the gauged four fermi model is
    $G=1,\beta=1$ for both mesons.  For the {}'t~Hooft model
    $\beta=1$ also.  For the Gross--Neveu model,  in the
    lightest meson case, the meson mass was chosen to agree with
    that of the gauged four fermi model.  In the next lightest
    meson case, $\mu^2/\beta=3$ was chosen.
    }
  \label{fig:wf}
  \end{center}
\end{figure}
As in the {}'t~Hooft model, the fermions are confined and the
following Regge--type behavior is observed for the highly
massive mesons, similarly to the {}'t~Hooft model:
\begin{equation}
  \label{tgn-regge}
  \mu^2\sim\pi^2 k\qquad k\gg1,\ \beta
\end{equation}
This Regge behavior is shown in \figno{regge}.
We may understand the behavior of the spectrum in the various
limits of the model as follows: when we turn off the
Gross--Neveu coupling $a^2$, the spectrum reduces to that of the
{}'t~ Hooft model.  As we take the gauge coupling to zero, which
effectively is the limit $\beta\rightarrow\infty$, the splitting
between the higher levels disappear.  We have explicitly checked
that the mass of the lightest meson approaches to the
Gross--Neveu model value plotted in \figno{gn-spectrum} in this
limit.  For the higher levels, $\mu^2$ approaches $4\beta$ in
this Gross--Neveu limit.  The chiral limit may be identified as
the limit $G\ ({\rm or}\ a^2N)\rightarrow\infty$, similarly to
the Gross--Neveu model case.  In the spectrum, mass of the meson
states decrease as we approach the chiral limit and it is clear
that $\mu^2\rightarrow0$ for the lightest meson bound state in
the limit $G\rightarrow\infty$.  When the limit
$\beta\rightarrow\infty$ is taken in addition, it may be
explicitly checked that the next lightest meson satisfies
$\mu^2/M^2\rightarrow4$ corresponding the the $\sigma$ mass in
the Gross--Neveu model.

As $\beta$ becomes large, the meson masses behave linearly with
the quark masses, as is expected from the naive constituent
quark picture. This picture is supposedly valid for highly
massive quarks.  The lightest meson behaves in a qualitatively
different manner from the other meson states in the theory.
This is a feature of the gauged four fermi model; such a
behavior does {\it not} occur in the {}'t~Hooft model.  The
Gross--Neveu coupling affects the lightest meson state
relatively more than the other meson states.  This disparate
behavior is a necessary consequence of the Gross--Neveu limit
where $\mu^2/\beta$ of $\chi$ and the other mesons approach the
corresponding value $\tmu^2$ in the Gross--Neveu model and four.
In the {}'t~Hooft model, $\mu^2/\beta$ for the lightest meson
also approaches four for large $\beta$.
\section{Summary and discussions}
\label{sec:disc}
In this work, we have solved the meson sector of the gauged four
fermi model in the large $N$ limit.  We provided detailed
prescriptions for solving the model systematically which should
be useful for further work involving this class of models.  We
also determined the physically consistent region in the gauged
four fermi model and analyzed the model there.  However, it is
possible that the regions with tachyons correspond to other
phases of the model that is inaccessible to our current methods.

Both the Gross--Neveu model and the {}'t~ Hooft model have been
used extensively in the literature to understand the physical
behavior of real systems, such as QCD.  A model that combines
the two models should be quite useful for understanding the
dynamics of various field theories.  In one direction, the four
fermi coupling has been used to model strong interaction
dynamics involving dynamical symmetry breaking for quite some
time \cite{t-gn}.  By interpolating between these two models, we
intend to shed more light on the relation between the physical
behavior of these two theories.  Furthermore, when dynamical
symmetry breaking scales are widely separated, in the
intermediate energy range, the theory effectively becomes a
gauged four fermi model. Such situations can occur quite
generally where the lower energy scale is the QCD scale or some
technicolor scale.  These types of theories are of
phenomenological interest and have been studied actively, for
instance, in the so called ``top quark condensation'' models
\cite{topc,t-gn}.  Admittedly, the model we studied is a $(1+1)$
dimensional toy model version of such theories. Historically,
however, $(1+1)$ dimensional theories have played an important
role in understanding of the corresponding higher dimensional
theories and we hope this will also be true in the future.

\bigskip

\noindent{\bf Acknowledgments: }   We would like to thank
Tomoyasu Ichihara for his collaboration during the early stages
of this work.  We would also like to thank K.~Itakura and
H. ~Sonoda for discussions.

\bigskip

\noindent {\large\bf Appendix: A brief note on the convergence of numerical
  solutions} \bigskip

The convergence of the numerical solutions depends on the
parameters $(G,\beta)$.  When $\beta\gtrsim1$, it is relatively
easy to achieve a relative accuracy of $\sim10^{-4}$ in the
meson mass at least using both the variational method (dimension
$\sim10$) and Multhopp's method (dimension $\sim400$).  For
$\beta\lesssim1$, more effort is needed to achieve the same
level of convergence.  The difficulties in the variational
method using polynomials of the momentum fraction (section
\ref{sec:variational}) arise because of the round off errors
since the eigenvalues in the normalization matrix
\eqnn{norm-elems}\ tend to become small.  Analytically choosing
an orthonormal basis or using some other basis appropriate for
the parameter region in question might alleviate this problem.
In Multhopp's method (section \ref{sec:multhopp}), the
limitations arise due to the necessary computational time when
using larger space of functions.  Choice of $g_m(\theta)$ may
speed up the convergence process in some parameter regions.

\begin{figure}[htb]
  \begin{center}
    \leavevmode
    \epsfxsize=\hsize\epsfbox{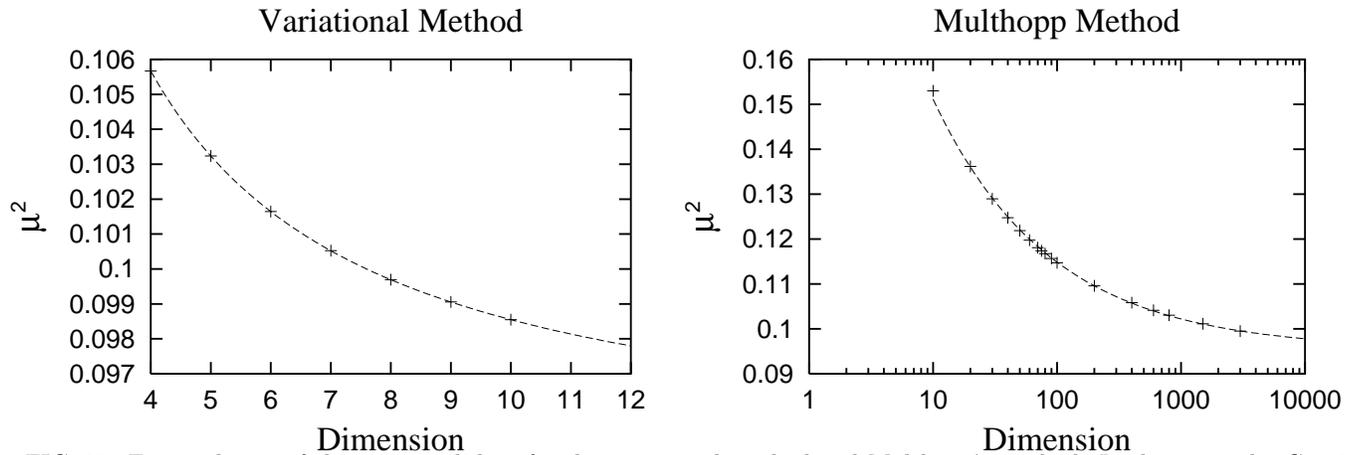}
    \caption{Extrapolation of the numerical data for the
    variational method and Multhopp's method.  
    In this example,
    $G=1$ and $\beta=0.1$ and $\mu^2=0.095\pm0.002$.} 
    \label{fig:convergence}
  \end{center}
\end{figure}

In the parameter regions where the convergence is slow,
extrapolation in the data can be effective.  We have found that
trial functions of the type $x+aK^b$ fits the data quite well.
Here $x$ is the extrapolated value, $K$ is the dimension of the
space span by the basis and $a,b$ are parameters.  Extrapolation
can sometimes be misleading so checks on the results are
desirable.  In our case, we compare the extrapolation values
from both the variational method and Multhopp's method and we
confirm that they are consistent within the errors of the fit.
An example of such an extrapolation is shown in
\figno{convergence}.

\bigskip

\end{document}